\documentclass[10pt, conference]{IEEEtran}
\usepackage{tikz}
\newcommand{\circled}[1]{\tikz[baseline=(char.base)]{\node[shape=circle,draw,inner sep=1pt] (char) {#1};}}
\usepackage{amsmath,amssymb,amsfonts}
\usepackage{graphicx}
\usepackage{textcomp}
\usepackage{diagbox}
\usepackage{booktabs}
\usepackage{colortbl}
\usepackage{multirow}
\usepackage{tabularx}
\usepackage{soul}
\usepackage{wrapfig}

\usepackage{times}
\usepackage{pifont} 
\usepackage{subfigure}
\usepackage{url}
\usepackage{breakurl}
\usepackage{paralist}
\usepackage{booktabs}
\usepackage{listings}
\usepackage{graphicx}
\usepackage{cite}
\usepackage{enumitem}
\usepackage{adjustbox}
\newcommand{\comment}[1]{}

\usepackage[caption=false]{subfig}
\usepackage[font=footnotesize,labelfont=bf]{caption}
\usepackage[normalem]{ulem}
\usepackage{algorithm}
\usepackage[noend]{algpseudocode}
\algtext*{EndWhile}
\algtext*{EndIf}
\algtext*{EndFor}
\algnewcommand\algorithmicforeach{\textbf{for each}}
\algdef{S}[FOR]{ForEach}[1]{\algorithmicforeach\ #1\ \algorithmicdo}
\usepackage{adjustbox}
\usepackage{xcolor}
\definecolor{myblue}{HTML}{93CDDD}
\definecolor{mygreen}{HTML}{17891B}
\definecolor{myyellow}{HTML}{DCC60F}
\definecolor{myorange}{HTML}{EB801B}
\usepackage{footnote}

\ifCLASSINFOpdf
\else
\fi
\begin{document}
%
\title{SoK: Securing the Final Frontier for Cybersecurity in Space-Based Infrastructure}

\author{Nafisa Anjum, Tasnuva Farheen\\Division of Computer Science and Engineering, Louisiana State University}
\maketitle

\begin{abstract}
With the advent of modern technology, critical infrastructure, communications, and national security depend increasingly on space-based assets. These assets, along with associated assets like data relay systems and ground stations, are, therefore, in serious danger of cyberattacks. Strong security defenses are essential to ensure data integrity, maintain secure operations, and protect assets in space and on the ground against various threats. Previous research has found discrete vulnerabilities in space systems and suggested specific solutions to address them. Such research has yielded valuable insights, but lacks a thorough examination of space cyberattack vectors and a rigorous assessment of the efficacy of mitigation techniques. This study tackles this issue by taking a comprehensive approach to analyze the range of possible space cyber-attack vectors, which include ground, space, satellite, and satellite constellations. In order to address the particular threats, the study also assesses the efficacy of mitigation measures that are linked with space infrastructures and proposes a Risk Scoring Framework. Based on the analysis, this paper identifies potential research challenges for developing and testing cutting-edge technology solutions, encouraging robust cybersecurity measures needed in space.  
\end{abstract}



%
\IEEEpeerreviewmaketitle

\section{Introduction}
Lyndon Johnson, a US senator at the time, stated in 1958 that commanding space infrastructure would equate to commanding the entire globe ~\cite{pavur2019cyber}. From engineering to the natural sciences, the study of space has significantly advanced technology and increased the scientific knowledge of humanity.  Additionally, it has improved our everyday lives in a number of ways; the European Space Agency (ESA) ~\cite{esa} claims that for every euro invested in the space sector, six euros are returned to the economy. Until lately, governmental support was associated with space since the space business was unappealing to corporations due to its large upfront costs and significant obstacles. These days, advances in satellite communications (Satcoms) technologies present special prospects for space research and development in the future ~\cite{huang2020recent, kodheli2020satellite}. Modern society heavily relies on space systems for various critical functions, including communication, navigation, weather forecasting, and national security. Satellites enable global internet access, GPS services, and real-time data transmission, supporting transportation, finance, and emergency response industries. This growing dependence on space infrastructure underscores the need to ensure its reliability, security, and resilience against both natural disruptions and cyber threats.

As our reliance on space-based systems grows, ensuring their security becomes increasingly critical. However, several challenges hinder this effort ~\cite{pravzak2021space}. Notably, there have been incidents where adversaries gained unauthorized access to mission-critical systems, such as the 2011 attack on NASA's Jet Propulsion Laboratory~(JPL), where attackers gained full control over mission-critical systems ~\cite{martin2012nasa}. In 2019, the U.S. Department of Homeland Security~(DHS) identified several incidents of GPS signal interference, which were suspected to involve state-sponsored actors. Such disruptions pose significant risks to both civilian and military operations that rely on precise satellite navigation, including transportation, logistics, and the guidance of precision munitions ~\cite{westbrook2019global}. A more severe incident occurred in 2022 when the KA-SAT satellite network, managed by Viasat, was targeted in a cyberattack that disrupted internet services across Europe. The attack, which focused on the satellite’s ground-based infrastructure, disabled modems and left tens of thousands of users—including military units—without satellite communication just before Russia's invasion of Ukraine ~\cite{halans2022viasat}. Despite these events, the cybersecurity landscape of space infrastructure—including its threats, vulnerabilities, and associated risks—remains under-explored. The widespread use of commercial off-the-shelf (COTS) components, the absence of comprehensive threat modeling, and the rise in cyber incidents underscore the necessity for a systematic review of existing research. Such an analysis would guide future cybersecurity strategies and solutions tailored for space infrastructure, aiding in informed decision-making, and drawing well-founded conclusions ~\cite{corallo2021understanding}.

Inspired by the above discussion, we present a thorough examination of cybersecurity for space by combining information from several disparate sources (both scientific and grey literature) that include several cyberattack and defense strategies that are currently in use. Furthermore, we propose a Risk Scoring Framework for assessing risk and adopting a proper mitigation strategy. 
To summarize, the following contributions are made in this work:
\begin{itemize}
    \item We offer a systematic evaluation of the current research comprising of 96 relevant publications.
    \item Combining the existing attack scenarios and differentiating between threats, we provide a thorough taxonomy.
    \item By analyzing the current gaps between threats and countermeasures, we identify important open challenges for future researchers.
    \item We model a framework for assigning a risk score to the cyberthreats in space assets and suggest an appropriate mitigation priority.
\end{itemize}
The rest of this paper is structured as follows: The research methodology is outlined in Section 2, while the detailed threat landscape and attack taxonomy  are provided in Section 3.  The countermeasures and their shortcomings are discussed in Section 4. Subsequently., a cybersecurity risk assessment model is proposed in Section 5. Based on the works analysed, recommendations and challenges for further research are provided in Section 6 and finally, Section 7 concludes the paper.
\vspace{-0.2in}
\section{Background}
The interconnectedness and unique operational challenges make the ground and space stations, satellites and satellite constellations high-value targets for cyber adversaries, underlining the critical need for robust, tailored cybersecurity strategies that span the entire space ecosystem. This section provides a review on how each of these segments work and why it is needed to protect them against cyber attacks.
\subsection{Ground Station}
Ground stations are terrestrial facilities equipped with antennas, signal processing hardware, and communication networks. They serve as the primary interface between space assets and Earth-based operations. These facilities receive telemetry data, command instructions, and other critical communications from satellites or space platforms. In many cases, they also process, analyze, and distribute data to end users and mission control centers. Ground stations may be integrated into broader networks that support real-time monitoring, control functions, and data storage.
\newline Why Cybersecurity Protection Is Essential:
\begin{itemize}
    \item Command and Control Vulnerabilities: Since ground stations send commands to satellites, any compromise can allow an adversary to issue unauthorized instructions, potentially hijacking or disabling space assets~\cite{lightman2022satellite}.
    \item Data Integrity and Confidentiality: Ground stations handle sensitive operational data, including scientific measurements, navigational information, and strategic communications~\cite{manulis2021cyber}. Cyberattacks could lead to data manipulation, leakage, or even denial of service, which might disrupt critical infrastructure.
\item Interconnected Networks: With many ground stations linked through public or private networks, a breach in one facility can propagate to others, amplifying risks across the entire space ecosystem.
\item Regulatory and Operational Impact: Given their role in national and commercial operations (e.g., weather forecasting, defense, communication), securing these systems is paramount to prevent cascading impacts on broader critical infrastructures.
\end{itemize}

\subsection{Space Station}
Space stations, such as the International Space Station (ISS), are habitable artificial satellites that serve as research laboratories, living quarters, and operational platforms in orbit. They provide a unique environment for scientific experiments, technology demonstrations, and even international cooperation. These platforms maintain continuous communication with Earth through dedicated ground stations and onboard communication systems, often utilizing complex internal networks to coordinate various systems like life support, navigation, and research instrumentation.
\newline Why Cybersecurity Protection Is Essential:
\begin{itemize}
    \item 
Mission-Critical Operations: Space stations support critical scientific and research activities, along with supporting international collaboration and long-duration human spaceflight. A successful cyberattack could compromise crew safety, disrupt experiments, or impair the station’s operational integrity.
\item Complex Networked Systems: The intricate networks onboard and the continuous link with Earth expose space stations to risks such as unauthorized access, data tampering, or malware propagation~\cite{botezatu2023attempted}.
\item Resource Constraints: Like satellites, space stations have limited onboard computational resources and must balance performance with security. This constraint makes it challenging to deploy advanced security solutions, which in turn heightens the risk of exploitation.
\item Interdependency with Ground Infrastructure: The reliance on external ground stations for updates, command, and data exchange means that vulnerabilities in either domain could have reciprocal adverse effects.
\end{itemize}

\subsection{Satellite}

Satellites are unmanned spacecraft that perform a variety of functions such as communication, remote sensing, navigation, and scientific observation. They are engineered with specialized hardware and software designed to operate in the harsh conditions of space. Typically, a satellite includes subsystems for power generation, communication, attitude control, and data processing. Once launched, satellites operate semi-autonomously but remain reliant on ground stations for command inputs and telemetry data exchange.
\newline Why Cybersecurity Protection Is Essential:
\begin{itemize}
    \item 
Limited Update Capability: Once deployed, satellites are difficult to physically access for repairs or updates. This makes pre-launch security measures and robust onboard defense mechanisms critical to withstand cyber threats over their operational life.
\item Communication Reliance: The bidirectional communication with ground stations exposes satellites to risks like signal interception, replay attacks, and unauthorized command injection, which can compromise satellite functionality or lead to operational disruption.
\item Critical Service Delivery: Satellites underpin vital services including global communications, weather forecasting, and navigation. Any successful cyber intrusion can disrupt these services, with wide-ranging economic and security implications.
\item Harsh Operational Environment: The unique space environment—with high radiation levels, temperature extremes, and isolation—complicates the implementation of conventional cybersecurity measures, necessitating tailored approaches that account for these constraints.
\end{itemize}

\subsection{Satellite Constellations}
Satellite constellations consist of large networks of satellites, typically in low Earth orbit (LEO), that work in unison to provide global coverage and enhanced service capabilities. By interconnecting hundreds or even thousands of satellites, these networks can offer robust services such as broadband internet, global positioning, and real-time data analytics. The architecture relies on sophisticated inter-satellite communications, coordinated orbital dynamics, and seamless integration with ground stations to achieve uninterrupted and high-capacity service delivery.
\newline Why Cybersecurity Protection Is Essential:
\begin{itemize}
    \item 
Cascade Risks: In a constellation, the failure or compromise of one satellite can have ripple effects, potentially disrupting the entire network’s operation. Cybersecurity measures must therefore account for interdependencies and implement safeguards to contain breaches.
\item Scalability Challenges: The sheer number of satellites and the dynamic nature of their orbits make constant monitoring and update of security protocols challenging. This complexity increases the likelihood of vulnerabilities going undetected.
\item High Service Impact: Constellations like those used for global internet services or navigation are integral to both commercial and defense sectors. A breach can impact millions of users, disrupt critical services, and have severe economic consequences.
\item Inter-satellite Communication Security: The reliance on secure, efficient inter-satellite links requires protocols that can manage issues like signal delay, intermittent connectivity, and the unique environmental challenges of space, making tailored cryptographic and authentication solutions a necessity.
\end{itemize}
\section{Methodology}

To identify and map the present shape of space cybersecurity and highlight areas for further research, the study used a structured, PRISMA~\cite{moher2009preferred}inspired literature review process. Prior to expanding the search by manually reviewing each paper's references, we conducted a search for relevant literature in online digital libraries.

\begin{table*}[ht]\scriptsize
\centering
\caption{Keyword mapping for literature search}
\label{tab:keyword-mapping}
\scriptsize
\renewcommand{\arraystretch}{1.5}
    \adjustbox{max width=\textwidth}{
  \begin{tabular}{p{3cm}p{4.5cm}p{3cm}}
  \toprule
  \textbf{Concept} & \textbf{Search String} & \textbf{Databases} \\
  \midrule
  space system security, satellite cybersecurity
    & \texttt{"space cybersecurity" OR "satellite cybersecurity"\newline OR "space system security"}
    & IEEE Xplore, ACM DL, Scopus, NDSS Symposium search, USENIX database \\
  ground segment security, ground station protection
    & \texttt{"ground station security" OR "ground segment security"\newline OR "ground station protection"}
    & IEEE Xplore, SpringerLink \\
  encrypted satcom, satcom security
    & \texttt{"secure satellite communication" OR "encrypted satcom"\newline OR "satcom security"}
    & USENIX database, ACM DL \\
  space cyber-physical, cyber physical attacks
    & \texttt{"space cyber-physical" OR "cyber physical attacks"\newline OR "cyber-physical attacks"}
    & IEEE Xplore, NDSS Symposium search \\
  risk scoring, risk quantification
    & \texttt{"cyber-risk assessment" OR "risk scoring"\newline OR "risk quantification"}
    & Scopus, USENIX database \\
  new space cybersecurity, COTS satellite security
    & \texttt{"commercial space operations" OR "new space cybersecurity"\newline OR "COTS satellite security"}
    & ACM DL, IEEE Xplore, CISA website \\
  satellite hacking incident, space cyberattack case
    & \texttt{"satellite hacking incident" OR "space cyberattack case"\newline OR "jamming case study"}
    & Google Scholar, C4ADS, NASA OIG \\
  \bottomrule
  \end{tabular}%
}
\end{table*}
\subsection{Identification}Initially, we conducted a search using the keywords in a number of well-known online digital libraries and proceedings. The detailed bibliographic databases along with boolean search strings are represented in Table ~\ref{tab:keyword-mapping}. We supplemented with targeted pulls from Agency websites (ESA, NASA OIG, CISA), conference proceedings not indexed above (IAC, AIAA), and high-profile incident repositories like Center for Advanced Defense Studies (C4ADS chronologies).

\subsection{Screening}For screening purpose, we used both scientific and grey literature.Literature based on the scientific method, which draws conclusions from evidence, is what we refer to as scientific literature. It develops theories and hypotheses based on earlier research while making sure to properly credit the authors and resources utilized. For this query, the keywords from Table ~\ref{tab:keyword-mapping} were utilized. Alternately, literature with constrained distribution—that is, not found in academic publishing libraries—is referred to as grey literature.  White papers, technical reports, policy documents, and unpublished reports are all included. The search returned 1,248 records spanning from the year 2003 to 2024. We removed 173 duplicates, yielding 1,075 unique records. Subsequently, inclusion and exclusion principles were applied:
\begin{itemize}
    \item \textbf{Include-} studies addressing cyber threats, vulnerabilities, attacks, defenses, or risk quantification explicitly for space assets (such as ground stations, satellites, constellations, space stations).
    \item \textbf{Exclude-}purely terrestrial/Industrial Control Systems (ICS) works, non cyber topics, and non English publications.
\end{itemize}
The screening resulted in 312 relevant records. 

\subsection{Eligibility}We retrieved 312 full texts and applied detailed eligibility criteria:
\begin{itemize}
    \item Explicit linkage to one (or more) of the four segments (ground, space station, satellite, constellation).
    \item Contains empirical data, simulation/analytical evaluation, or detailed conceptual framework.
\item	Provides sufficient methodological detail for coding.
\end{itemize}
We excluded 238 papers (e.g., lacking space focus, insufficient methodological rigor, or inaccessible full text).
74 studies met these criterias. To capture emerging work not yet indexed, we performed backward snowballing and forward citation tracking on all 74 references via Google Scholar thus adding 22 more relevant articles. Hence, the final scope of the study included 96 relevant records in the corpus.

\subsection{Evaluation and Modeling} In order to extract crucial security-relevant information, including attack pathways, models and taxonomy, target components, we first read and examined each document in this step.  Second, we matched every attack and response strategy to the appropriate space segment, threat, or mitigation category. The outcome of this modeling and data evaluation was the identification of the fundamental components for our initial draft's research questions.The subsequent inquiries for research were established:

\noindent\circled{1}  What are the prevalent cyberattacks in current space infrastructure?

\noindent\circled{2} What types of faults can existing countermeasures address?

\noindent\circled{3}  What research gaps do exist in between current mitigation techniques and real world cyberattacks?

\noindent\circled{4}  How to assess cyber-risk to take prompt action and initiate recovery?

Following these questions, the analyzed works examine many cybersecurity topics, including threats, risks, countermeasures, exisitng gaps, regulations, and requirements for space cybersecurity.

\section{Threat Landscape and Attack Taxonomy}
\begin{figure*}[ht]
    \centering
    \includegraphics[width=0.65\textwidth]{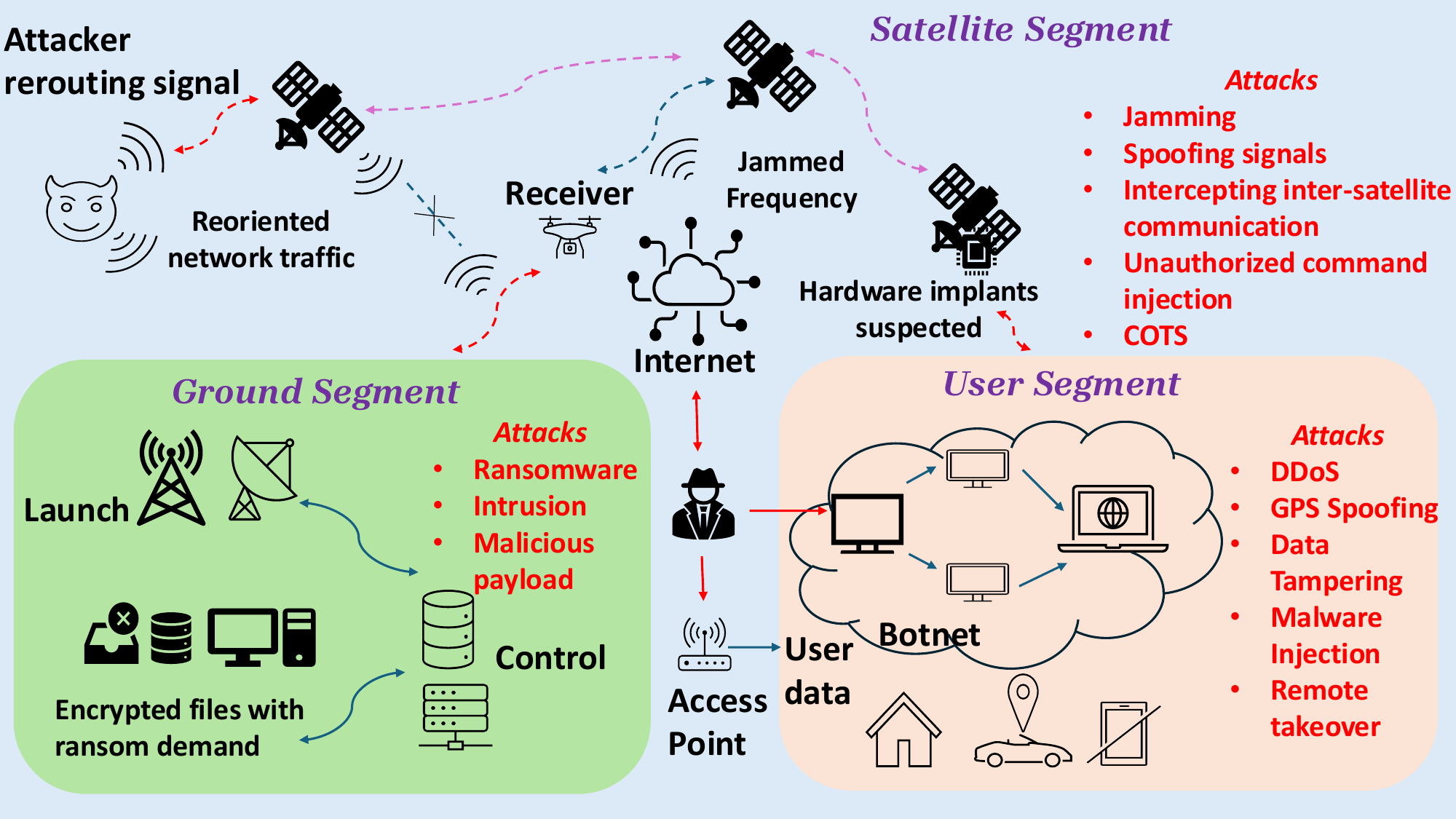}
    \caption{Space infrastructure and attack taxonomy}
    \label{fig:attacks}
\end{figure*}

Space cybersecurity threats have expanded due to technological improvements, multistakeholder fragmentation, and higher investment.  The force field of cybersecurity issues therefore encircles future missions.  The ground segment, satellite segment, space segment and satellite constellations are all possible cyberattack vectors, as shown in Figure~\ref{fig:attacks}. The process of "meta-synthesis," which is the comprehensive examination and integration of results from qualitative literature, is used to derive these four fragments~\cite{walsh2005meta}.  The subsequent subsections provide a detailed discussion of each fragment.

\subsection{Attacks on Ground Station}
Attacks on the ground segment (GS) are particularly alarming considering satellites are inherently susceptible to cyberattacks.  Satellites and broader satellite services are overseen here.Attackers aiming for the ground station can illegally obtain access and control by taking advantage of vulnerabilities. They are capable of carrying out a number of attacks, such as manipulation of data, DoS attacks, malware attacks, cloud-based attacks, and illegal access ~\cite{lightman2022satellite}. According to the STRIDE paradigm, attackers may try to access GSs without authorization, which is classified as an elevation of privilege threat ~\cite{atmaca2022challenges}. Cybercriminals can alter satellite control systems and carry out forbidden orders by breaching operator or administrative accounts ~\cite{hudaib2016satellite}. Moreover, GSs and the general integrity of satellite communications (Satcomm) systems are seriously threatened by data alteration attempts ~\cite{hayes2023cyber}. Technical glitches or interruptions in GS operations can also result in purposeful or unintentional data alterations ~\cite{manulis2021cyber}. The study in~\cite{hamill2024threats} presents empirical incident data on GS compromises, categorizes attack vectors (network, physical, supply chain), and evaluates mitigation efficacy in live testbeds. Furthermore, the telemetry data sent from the satellite to the GS may be the target of data alteration attempts. Important details regarding the satellite's condition, efficiency, and health are included in telemetry data ~\cite{moustafa2022dfsat}. Attackers may conceal system flaws, display erroneous measurements, or interfere with the GS's capacity to precisely track the satellite's status by manipulating this data ~\cite{manulis2021cyber}. Without technically targeting the systems, physical attacks such illegal access to ground stations and other tangible assets can shut down the ground station, endangering the space mission's ability to function and taking control of the space assets and their operations.  The International Space Station's command and control algorithms were compromised after an unencrypted notebook computer was stolen, according to a NASA report ~\cite{martin2012nasa}.  Two NASA satellites were taken over by ground stations in 2007 and 2008 ~\cite{zatti2017protection,bardin2025satellite}. Additionally, like any other computer system, the ground element of Satcoms is subject to imminent risk from software vulnerabilities ~\cite{falco2019cybersecurity}. Besides, Denial of service (DoS) and Distributed denial of service (DDos) are extremely disruptive cyberattacks in the GS which render a network or system inoperable by flooding it with excessive traffic, making it inaccessible.
to authorized users ~\cite{tedeschi2022satellite}.It becomes extremely difficult to identify the attack's origins and separate malicious traffic from original inquiries ~\cite{usman2020mitigating}.
\vspace{-0.1in}
\subsection{Attacks on Space Station}
The satellites and their onboard subsystems are part of the space segment, which is susceptible to a number of vulnerabilities that could jeopardize security measures and operations ~\cite{he2019security}.Cyberattacks that target the hardware and software components of satellites explicitly can affect this sector of Satcoms systems ~\cite{sawikspace}.To accomplish their destructive goals, for instance, disrupting communication channels and stealing sensitive data from satellites, cybercriminals may utilize a variety of strategies, such as malware ~\cite{konstantinou2019hardware}, DoS assaults ~\cite{he2019security}, and other cyberthreats. Additionally, using easily available and affordable Commercial Off-The-Shelf (COTS) hardware and software in space infrastructures may result in the development of new points of vulnerability ~\cite{nussbaum2020cybersecurity}.Additionally, unpatched or out-of-date software may put the space section at jeopardy making it susceptible to exploitation ~\cite{manulis2021cyber}. Furthermore, the communication between space station and the ground segment is obstructed due to link jamming that results in unreliable information ~\cite{de2013eutelsat}. 
\vspace{-0.1in}
\subsection{Attacks on Satellites}
The attacks on satellites have been analyzed in several literature. Spoofing is one of the most common attacks where the goal is to intercept, modify, and retransmit a communication signal in order to deceive the recipient into believing it came from the designated sender.  By posing as an authorized user and issuing fictitious commands, spoofing attacks on satellites entail seizing control of a space communication system and causing the spacecraft to fail or malfunction during its mission ~\cite{boschetti2022space, amin2023stochastic}.By interfering with the radio transmission that satellites utilize to receive commands, the jamming attack may have an impact on a satellite's regular operation ~\cite{falco2018job}.  A number of malicious individuals in space, including nation-states, professional or amateur hackers, organized crime, and insiders, have been taken into consideration while analyzing the jamming danger, which is covered in ~\cite{vollmer2021natos}. The tampering threat, where, by obtaining illegal access to the satellite systems, an adversary can add, remove, or alter files have been discussed in several works ~\cite{jha2022safeguarding, fleming2023securing}. The STRIDE and DREAD techniques are used to examine DoS threats in satellites after an exhaustive examination ~\cite{atmaca2022challenges}. A brief discussion of DoS in relatively small satellites is given in ~\cite{saha2019ensuring} , taking into account network, software, and hardware vulnerabilities.
\vspace{-0.1in}
\subsection{Attacks on Satellite Constellations}
In contemporary Satcoms systems, satellite constellations—which are made up of several interconnected satellites—are being used more and more to offer continuous, worldwide coverage ~\cite{graczyk2021sanctuary}.  Nevertheless, this interconnectedness creates vulnerabilities that mostly fit within the DoS and tampering categories of the STRIDE paradigm ~\cite{tedeschi2022satellite}.A specific satellite may be the target of an attacker who compromises its control systems or communication lines. This compromised spacecraft may be used as a springboard for additional cyberattacks on other satellites in the constellation once command has been established ~\cite{manulis2021cyber}.Attackers might take advantage of satellite constellation tampering vulnerabilities ~\cite{zhang2022security}.  In addition to endangering the compromised satellite, this interference may have repercussions that could undermine the constellation's ability to function as a whole. Moreover, Man-In-The-Middle (MitM) attacks also compromise the reliability within the communication channel ~\cite{aghayev2024i2s}. Furthermore, currently, a 1 kg "Cube Satellite" that is fully built costs \$16,000 ~\cite{pavur2020sok}. Such availability of pre-built satellite flight gear lowers procurement costs, enabling New Space firms to assume greater commercial and technical risks from COTS satellite components.
\vspace{-0.1in}
\subsection{Systematic Comparison of Cyberthreats}
To systematically analyze cyber threats in space infrastructure, the following tables compare various aspects such as differences from traditional cybersecurity, attack vectors, impact severity and effectiveness of mitigation strategies. These comparisons highlight how space systems face unique challenges due to their remote and autonomous nature, reliance on RF communication, and limited physical security measures.

\begin{table}[htbp]
    \centering
    \caption{Comparison of Space Cybersecurity with Traditional Cybersecurity}
    \label{tab:traditional_vs_space}
    \renewcommand{\arraystretch}{1.2}
    \adjustbox{max width=\columnwidth}{
    \begin{tabular}{@{}p{2.6cm}p{2.9cm}p{3.3cm}@{}}
        \toprule
        \textbf{Feature} & \textbf{Traditional Cybersecurity} & \textbf{Space Cybersecurity} \\
        \midrule
        Accessibility & Easy to patch remotely & Limited update capability \\
        Latency & Low & High due to long distances \\
        Encryption Usage & Standardized & Often outdated or absent \\
        Physical Security & Possible local access & Almost impossible due to orbital location \\
        Attack Surface & Primarily network-based & Includes RF, supply chain, AI-based vulnerabilities \\
        \bottomrule
    \end{tabular}}
\end{table}

Cybersecurity strategies for space must be adapted to ensure resilience against both conventional and space-specific threats. Hence, Table~\ref{tab:traditional_vs_space} analyses the differnces between traditional and space domains for cybersecurity in terms of five prime features. Due to low accessibility and outdated software or encryption techniques, space assets are often compromised. Furthermore, direct intervention (e.g., fixing hardware issues, installing physical security measures) is impractical or prohibitively expensive for these assets. Once deployed, hardware remains largely unmodifiable unless designed with self-repairing mechanisms or redundancy. Apart from coventional risks, modern AI-driven space infrastructures may unintentionally reveal sensitive information due to adversarial attacks which make them more vulnerable.

\begin{table*}[htbp]
    \centering
    \caption{Comparison of Space Cyber Threats by Attack Vector}
    \label{tab:attack_vector}
    \renewcommand{\arraystretch}{1.4}
    \adjustbox{max width=\textwidth}{
    \begin{tabular}{@{}llp{4cm}p{5cm}p{4cm}p{4cm}@{}}
        \toprule
        \textbf{Attack Vector} & \textbf{Attack Type} & \textbf{Targeted Systems} & \textbf{Technical Method} & \textbf{Impact} & \textbf{Real-World Example} \\
        \midrule
        \multirow{2}{*}{\textbf{Radio Frequency (RF) Attacks}} & Jamming & Satellites, Ground Stations & High-power RF signals disrupt legitimate transmissions & Loss of control, data transmission failure & Russian GPS jamming during military operations~\cite{GPSJamming2024} \\
        & Spoofing & Navigation Satellites (GPS, GNSS), Telemetry Links & Fake signals injected to alter positioning and telemetry data & False location/navigation, misinformation & GPS spoofing in Black Sea misleading ships~\cite{C4ADS2022} \\
        \midrule
        \multirow{2}{*}{\textbf{Network-Based Attacks}} & DoS/DDoS & Ground Control, Space-Based Networks (Starlink) & Overloading networks with excessive requests & Ground control failure, degraded satellite communications & Suspected cyberattacks on satellite ISPs~\cite{ViasatAttack2022} \\
        & Man-in-the-Middle (MitM) & Ground-Satellite Uplinks, Cross-Satellite Communication & Intercepting and modifying transmitted data & Data theft, unauthorized command injection & Theorized interception of military satellite data~\cite{cernat2024impact} \\
        \midrule
        \multirow{2}{*}{\textbf{Malware and Exploits}} & Ransomware & Ground Stations, ISS, Satellites & Encrypting files or system control with ransom demand & Locking out mission control, loss of critical data & ISS laptop malware infection~\cite{Wikipedia2008Malware} \\
        & Firmware Exploits & Satellite OS, Onboard Flight Computers & Exploiting software vulnerabilities in satellite firmware & Unauthorized control, long-term compromise & Theorized Chinese satellite firmware backdoors~\cite{CISA2023} \\
        \midrule
        \multirow{2}{*}{\textbf{Physical and Supply Chain Attacks}} & Hardware Backdoors & Satellite Processors, Navigation Chips & Pre-installed malicious circuits or logic bombs & Persistent access, undetectable long-term exploitation & Suspected hardware implants in defense satellites~\cite{USSF2024} \\
        & Insider Threats & Space Station Networks, Ground Ops & Rogue actors leaking or manipulating critical data & Espionage, sabotage, misconfigurations leading to failure & NASA employee accused of leaking classified data~\cite{NASA_OIG} \\
        \bottomrule
    \end{tabular}}
\end{table*}

Given the unique challenges of securing space systems, a breakdown is necessary to understand the diversity of attack types and their consequences on satellite communications, ground control, and mission operations.As such, Table~\ref{tab:attack_vector} provides a structured comparison of different cyber threats in space infrastructure, categorizing them based on the attack vector, targeted systems, technical methods, impact, and real-world examples. Attack vector includes Radio Frequency (RF) Attacks, Network-Based Attacks, Malware and Exploits, and Physical and Supply Chain Attacks whereas Attack Type specifies specific attack techniques under each attack vector. Navigation satellites (GPS, GNSS), ground stations, onboard flight computers, and even human-operated networks like the ISS are the components within the space infrastructure that are often affected. By analysing how the attack is performed and the impacts of a successful attack, preventative measures can be taken without delay. As space operations become more commercialized (e.g., SpaceX, OneWeb, and NASA Artemis) and integrated with AI, the risk of cyberattacks will only increase which has been observed from real world cases.

\begin{table*}[htbp]
    \centering
    \caption{Comparison of Cyber Threats by Impact Severity}
    \label{tab:impact_severity}
    \renewcommand{\arraystretch}{1.3}
    \adjustbox{max width=\textwidth}{
    \begin{tabular}{@{}llp{5cm}p{5cm}p{5cm}@{}}
        \toprule
        \textbf{Threat Type} & \textbf{Severity Level} & \textbf{Potential Consequences} & \textbf{Recoverability} & \textbf{Notable Case} \\
        \midrule
        Jamming & Medium & Temporary disruption of satellite communication & High – Frequency hopping mitigates this & GPS jamming in military zones \\
        Spoofing & High & False positioning, misinformation affecting operations & Medium – Cryptographic authentication helps & GPS spoofing misleading aircraft and naval vessels \\
        DoS/DDoS & Medium & Ground control and satellite communication failure & High – Network redundancy and AI-based filtering & Suspected cyberattack on Starlink~\cite{abdulla2024starlink} \\
        Firmware Exploits & High & Long-term system compromise, unauthorized access & Low – Requires remote firmware patching, difficult in orbit & Potential backdoor exploits in spacecraft firmware \\
        Ransomware & High & Loss of control over mission-critical systems & Medium – Backup systems and redundancy may mitigate & ISS laptop infected by USB-borne malware~\cite{Wikipedia2008Malware} \\
        Hardware Backdoors & Critical & Stealthy, long-term exploitation, data theft & Very Low – Requires physical hardware replacement & Theorized implanted vulnerabilities in military satellites \\
        \bottomrule
    \end{tabular}}
\end{table*}

Table~\ref{tab:impact_severity} has been analysed for understanding how cyber threats differ in their severity and recoverability in space environments as per NIST SP 800-30 guidelines ~\cite{scholl2023introduction}. High risk threats such as firmware exploits and hardware backdoors has very low recoverability and requires physical intervention in most cases. On the other hand, threats like jamming and DoS can be dealt with effectively.By categorizing cyber threats based on severity and recoverability, this table helps prioritize threats and enhance security measures.
\section{Existing Mitigation Strategies and Gap Analysis}
Securing space infrastructure requires a multi-layered cybersecurity approach due to the unique constraints of space systems, such as long mission lifespans, limited computational power, and remote operation. Below are the key countermeasures used to protect space assets from cyber threats.

\subsection{Communication Security}
Malicious actors use the communication link as their main and broadest attack surface, taking advantage of flaws in satellite communication protocols.
Certain characteristics, such as safe handover systems and anti-jamming approaches, are worth taking into account while developing protocols ~\cite{tedeschi2022satellite}.  Directive antenna technology, game theory/reinforcement learning, and spread spectrum may all provide a basis for anti-jamming strategies.  Inter-satellite link, flood, and cooperative routing are examples of secure routing systems that guarantee the confidentiality and integrity of data transfers.  In the dynamic environment of space missions, secure changeover systems that are based on inter-satellite, beam, and node mobile handovers further strengthen the robustness and dependability of communication networks. Anti- jamming and anti-spoofing strategies utilize spread spectrum techniques such as Frequency Hopping Spread Spectrum (FHSS), Direct Hopping Spread Spectrum (DHSS) ~\cite{kumar2022biometric,wang2021overview} along with directional antennas ~\cite{shvets2018antenna} and cryptographic authentication ~\cite{lu20195g}.~\cite{csiac2024implementing} details a case study on deploying quantum‑safe key distribution to ground stations, plus evaluation of VPN and Zero‑Trust overlays in operational SatCom networks.

Thus, putting strong encryption techniques into practice and using secure communication methods is essential to strengthen data reliability, communication between ground stations, spacecraft, and other mission elements. Weighing the trade-off between cybersecurity and quality of service (QoS) safety ~\cite{khan2021security}, communication protocols need to be considered carefully.  One of the QoS requirements is the delivery of data in real time at increased internet speeds, safe data transfer, and compatibility between in-space objects, as well as smooth end-to-end user interaction as well as space instruments.  Hence, cybersecurity requires a strong communication channel with seven key components: accessibility, robustness, integrity, confidentiality, dependability, and reliability.

\subsubsection{Gaps in Communication Security: }
Current space communication protocols, such as Secure Shell (SSH) and Space Communications Protocol Standards (SCPS), have shortcomings and possible risks ~\cite{hogie2005using}.  These include potential flaws in protocol implementation, difficulties managing keys, vulnerability to MitM attacks, and deficiencies in password-based authentication. Importantly, these procedures might not provide sufficient protection against social engineering attacks, illegal access, and insider threats.  This emphasizes how important it is to implement best practices and extra security measures in order to successfully reduce these complex threats.

\subsection{Secure Software $\&$ Firmware Protection}
Securing software and firmware in space systems is crucial to ensure mission integrity and resilience against cyber threats. Recent research has highlighted several approaches to enhance this security. For instance, ~\cite{scharnowski2023case} conducted a case study on fuzzing satellite firmware, emphasizing the importance of proactive vulnerability discovery in space systems.Additionally, the seL4 microkernel has been formally verified to ensure functional correctness, offering a robust foundation for secure software architectures in space applications ~\cite{Wikipedia_L4Microkernel}. Furthermore, the development of frameworks for secure firmware updates, such as the one proposed by ~\cite{falas2021modular}, provides modular end-to-end solutions to protect embedded systems from unauthorized modifications. These advancements collectively contribute to strengthening the cybersecurity posture of space infrastructure.

\vspace{-0.04in}
\subsubsection{Gaps in Secure Software $\&$ Firmware Protection:}
Satcoms systems still face a number of significant firmware and software upgrade issues ~\cite{he2019security}. The transmission of software and firmware updates can be slowed down and made more difficult by satellites' frequent struggles with limited bandwidth ~\cite{perez2019signal}. Remote locations of satellites can make it more difficult to identify and fix problems that may occur during or after an update ~\cite{ingols2017design}. Satellites are built to be extremely dependable and to function flawlessly over extended periods of time. It can be challenging to reduce the risks and potential points of failure that software and firmware changes can bring about ~\cite{sawik2023space}. Complex and customized software and firmware are frequently used by satellites, and they might not be compatible with the contemporary standards or technology ~\cite{mangan2022experiences}. 

\subsection{Access Control Security}
Implementing strong access control management solutions is a crucial step in reducing security threats inside the space system, given the wide range of stakeholders involved in missions.  To guarantee that only those with the proper authorization can access vital systems and data, strict authentication procedures and access control measures must be implemented.The implementation of Zero Trust Architecture (ZTA) ~\cite{stafford2020zero} has emerged as a pivotal strategy in this domain. Unlike traditional security models that rely on perimeter defenses, ZTA operates on the principle of "never trust, always verify," ensuring that every user and device is continuously authenticated and authorized before accessing resources. The Cybersecurity and Infrastructure Security Agency (CISA) underscores the significance of ZTA in space environments, highlighting its role in mitigating risks associated with credential compromises and unauthorized lateral movements within networks ~\cite{CISA_ZeroTrustSpace}. Furthermore, the adoption of blockchain technology offers promising advancements in decentralized access control for space systems. For instance, Xu et al.~\cite{Xu_2019} propose a blockchain-enabled strategy that enhances identity authentication and fine-grained access management, addressing challenges inherent in the decentralized and heterogeneous nature of space networks. A thorough analysis of blockchain's possible uses in the context of multi-sensor satellites has been conducted by de La Beaujardiere and Mital ~\cite{de2019blockchain}, adding to the growing debate about incorporating cutting-edge technologies to improve the functionality and security of space systems.

\textbf{Gaps in Access Control Security: }
Despite significant advancements, several gaps remain in securing space-based networks and access control mechanisms due to the unique constraints of space environments. While Zero Trust Architecture (ZTA) is gaining traction, its adaptation for resource-constrained space systems remains underexplored. In the case of efficient key management, current cryptographic key distribution methods lack flexibility and cannot be updated efficiently post-launch. Moreover, traditional Public Key Infrastructure (PKI) is difficult to implement due to high latency and lack of centralized authorities in deep space networks. Latency and bandwidth limitations in space communications make continuous authentication and access verification challenging. Additionally, Space networks lack dynamic security policies due to the rigidity of traditional hardware-based network architectures. Software-defined networking (SDN) could improve flexibility, but its security risks (e.g. compromised SDN controllers) remain understudied. Inter-satellite communications, furthermore, lack standardized security protocols for cross-vendor authentication. Satellite-to-ground station authentication relies heavily on pre-configured credentials, increasing the risk of credential theft or replay attacks.
\vspace{-0.04in}
\subsection{Intrusion Detection and Response Mechanisms}
Intrusion Detection and Prevention (IDP) systems are a collection of methods and resources intended to keep surveillance on and protect the different segments of space infrastructure against cyberattacks in the context of Satcoms cybersecurity ~\cite{cao2021blockchain, ashraf2022deep}. Using a signature database, signature-based detection ~\cite{shaikh2022advanced} efficiently blocks known threats by searching network traffic for particular patterns or signatures linked to known threats.Network traffic that exhibits odd or suspicious activity can be identified using anomaly-based detection ~\cite{obied2023deep, she2016intrusion}.  Machine learning (ML) techniques can help the system recognize anomalies and spot potential dangers by creating a baseline of typical behavior. Network traffic is monitored using network-based intrusion detection ~\cite{li2020distributed, uhongora2023deep} to find indications of infiltration and identify attacks directed at several networked devices.

\textbf{Gaps in Intrusion Detection and Response Mechanisms: }
Unlike terrestrial IDP, IDP systems for space face unique constraints, such as high latency, limited computational resources, and difficulty in real-time response. Existing IDP models (e.g., signature-based, anomaly-based, AI-driven IDP) lack universal compatibility across different satellite platforms, mega-constellations, and deep-space missions. Traditional IDP mechanisms require continuous network monitoring, which is challenging for satellites due to limited computational resources and power constraints. Furthermore, Most IDP solutions for space focus on detection only, but few offer automated mitigation (Intrusion Prevention Systems - IPS). AI/ML-based models that have been deployed for IDP recently are vulnerable to adversarial attacks, where attackers manipulate input data to evade detection. Space-Based Intrusion Detection and Prevention Systems (SIDPS) need more adaptive, autonomous, and lightweight architectures to function efficiently in space environments.

\subsection{Supply Chain and Hardware Security}
Several security techniques have been proposed to safeguard the supply chain and space hardware. Physical Unclonable Functions (PUFs) have been widely researched as a means to uniquely identify and authenticate ICs, ensuring that only verified components are used in critical satellite subsystems ~\cite{herder2014physical}. Additionally, side-channel analysis has been leveraged to detect hardware Trojans and anomalies in cryptographic operations, as demonstrated by Yang et al. ~\cite{yang2021side}. In order to address vulnerabilities in the supply chain due to COTS satellite components, blockchain-based supply chain tracking is emerging as a promising solution for ensuring provenance and traceability of space-grade components ~\cite{ramachandran2020towards}. Blockchain can provide tamper-resistant records of component sourcing, reducing the risks of counterfeit infiltration. Furthermore, AI-driven anomaly detection is being explored for real-time monitoring of satellite hardware integrity, with research suggesting the use of machine learning models to detect unauthorized modifications in firmware ~\cite{diana2024review}.

\textbf{Gaps in Supply Chain and Hardware Security: }
Despite these advancements, several research gaps remain in this domain. The effectiveness of hardware Trojan detection methods is still limited due to high false-positive rates and the difficulty of inspecting complex, nanoscale circuits post-manufacturing. Additionally, globalized supply chains mean that satellites often incorporate parts from multiple vendors, increasing the risk of supply chain attacks which cannot be traced completely yet.
\vspace{-0.1in}
\subsection{Standards and Regulations to Direct Secure Operations}
To provide consistent cybersecurity procedures throughout the space industry, adherence to international norms and guidelines is essential. There are a number of fundamental standards that offer an acceptable framework for protecting space systems.The Consultative Committee for Space Data Systems, or CCSDS, offers guidelines for protecting space mission procedures, particularly with relation to data transfer, network security, and encryption.  These guidelines provide a standard for developing safe systems in orbit ~\cite{CCSDS_350x1g3}. The standards for putting in place an information security management system (ISMS), which can be modified for space operations, are outlined in ISO 27001.  An important tool for protecting space-based assets is ISO 27001, which focuses on effectively handling confidential data ~\cite{botezatuspace}. The NIST SP 800-160 standard encourages the integration of security across the system development lifecycle and places a strong emphasis on systems security engineering.  Space enterprises may create more robust systems that can survive present and future cyberthreats by embracing a secure by-design attitude ~\cite{krumay2018evaluation}.  
\vspace{-0.1in}
\begin{table*}[htbp]
    \centering
    \caption{Risk Assessment Model Parameters}
    \label{tab:risk_score}
    \renewcommand{\arraystretch}{1.3}
    \adjustbox{max width=\textwidth}{
    \begin{tabular}{@{}llp{6cm}cc@{}}
        \toprule
        \textbf{Risk Factor} & \textbf{Symbol} & \textbf{Description} & \textbf{Weight (W)} & \textbf{Scale} \\
        \midrule
        Threat Likelihood & L & Probability of a cyberattack based on threat intelligence & 0.25 & 1--10 \\
        System Vulnerabilities & V & Number and severity of known vulnerabilities & 0.20 & 1--10 \\
        Attack Surface & A & Size and complexity of exposed interfaces (uplink, downlink) & 0.15 & 1--10 \\
        Impact of Attack & I & Consequences of an attack on mission success and safety & 0.30 & 1--10 \\
        Access Control Effectiveness & C & Strength of authentication, encryption, and privilege controls & 0.10 & 1--10 \\
        \bottomrule
    \end{tabular}}
\end{table*}
\section{Cybersecurity Risk Assessment for Space Infrastructure}
Space systems are increasingly vulnerable to cybersecurity threats that can compromise mission integrity, disrupt communications, and pose national security risks. A structured risk Scoring model helps in quantifying and prioritizing threats based on their severity and likelihood, ensuring efficient cybersecurity strategies. Risk analysts and inspectors deal with a lot of challenging issues pertaining to emerging cyber systems. These difficulties include the ever-evolving character of cyber systems due to technological advancements, their dispersion throughout the information, physical, and sociocognitive domains, and their intricate network architectures, which often consist of thousands of nodes. In order to get over some of the obstacles that cyber risk assessment faces, here we propose a Multi-Criteria Decision Making (MCDA) ~\cite{linkov2011multi} approach that quantifies cyber threats and vulnerabilities within the space infrastructure. This approach consists of two main steps: scaling key parameters as per threat characteristics, calculating risk score and taking immediate mitigation action interpreted from the score. The elements of the quantitative cybersecurity Risk Scoring Framework have been discussed here.

\noindent \underline{\textbf{ Key Risk Factors and Weighting:}}
The proposed structure in ~\cite{ganin2020multicriteria} is intended to evaluate a cyber system's risk using threats, vulnerabilities and consequences as the most significant criterias in order to choose the best remedial strategy. Expanding on their idea, the scoring mechanism developed here scores each risk factor from 1 (low risk) to 10 (high risk) in Table. \ref{tab:risk_score}. We began by assigning weights to the five risk criterions, namely, Threat Likelihood, System Vulnerabilities, Attack Surface, Impact of Attack and Access Control Effectiveness. These weights would be obtained from security specialists using established procedures ~\cite{buede2024engineering} in an empirical implementation of this paradigm, depending on the attributes of the cyber system.
The weights add up to 1.00 to ensure probabilistic balanced scoring. The values of the scale have been interpreted as per NIST SP 500-53 guidelines ~\cite{maclean2017nist} but are susceptible to change on the basis of threat likelihood. 

\noindent \underline{\textbf{Formula and Scoring Interpretation:}}
Subsequent to defining and quantifying the parameters, the overall cybersecurity risk score is calculated using a weighted sum in Equation. \ref{equ-riskscore}. Access Control (C) is subtracted from 10 because stronger controls reduce risk. 
\begin{equation}
\label{equ-riskscore}
\begin{split}
    \text{Risk Score} =\ & L \times W_L + V \times W_V + A \times W_A \\
    & + I \times W_I + (10 - C) \times W_C
\end{split}
\end{equation}
Using the formula, the risk score that is calculated is assigned Risk Levels from Low to Critical (Table. \ref{tab:score_interpret}). From the risk level, the required mitigation priority and appropriate action to be undertaken for the cyberattack can be determined.
\begin{table}[htbp]
    \centering
    \caption{Risk Scoring Interpretation}
    \label{tab:score_interpret}
    \renewcommand{\arraystretch}{1.5}
    \adjustbox{max width=\columnwidth}{
    \begin{tabular}{@{}clll@{}}
        \toprule
        \textbf{Score Range} & \textbf{Risk Level} & \textbf{Mitigation Priority} & \textbf{Actions} \\
        \midrule
        1 -- 3 & Low & Routine monitoring & Minimal action required \\
        4 -- 6 & Moderate & Implement preventive measures & Risk reduction recommended \\
        7 -- 8 & High & Immediate risk mitigation required & Urgent action needed \\
        9 -- 10 & Critical & Emergency response & Highest priority \\
        \bottomrule
    \end{tabular}}
\end{table}
An organization prioritizes outcomes and controls that can manage the risks with the most negative impacts and/or that are most cost-effective for their risk management results by using the principles outlined in NIST SP 800-30, Guide for Conducting Risk Assessments ~\cite{scholl2023introduction}. Based on the principles outlined by NIST, our Risk Scoring Framework can be utilized to interpret the required level of action for threat mitigation. 
For instance, consider a satellite with the following data for jamming or spoofing attack:

\noindent\circled{1} Likelihood (L): 8 (high probability of attack)

\noindent\circled{2} Vulnerabilities (V): 7 (moderate number of known weaknesses)
\noindent\circled{3} Attack Surface (A): 6 (moderate exposure through uplinks or downlink)
\noindent\circled{4} Impact (I): 9 (severe impact on mission success)
\noindent\circled{5} Access Control (C): 8 (strong encryption and authentication)

Calculation of risk score:
\begin{equation*}
\begin{split}
    \text{Risk Score} =\ & (8 \times 0.25) + (7 \times 0.20) + (6 \times 0.15) \\
    & + (9 \times 0.30) + (10 - 8) \times 0.10
\end{split}
\end{equation*}

Result: 7.2 (High Risk) — Immediate mitigation required.

Similarly, the risk score for a satellite with moderate likelihood and impact of Denial of Service attack on sensor can be estimated to have a Moderate risk score (around 6.5) where preventative measures need to be implemented. Hence, an organization could implement this cybersecurity Risk Scoring Framework procedures to assess and resolve potential security threats. 
\vspace{-0.1in}
\section{Open challenges for Cyberattacks in Space Systems}
In order to address various unresolved issues, this section discusses the primary outstanding difficulties that space systems face and suggests future avenues of research and development.  A space system's usability and the resource costs (such as energy, processing cycles, and memory) for the security mechanisms must be carefully weighed against the amount of security offered in order to deliver the level of service that users require.  In general, all of the prospective research avenues mentioned in this section fall within this balanced approach. 

Space assets are susceptible to attacks and vulnerabilities at every stage of a system's lifecycle.  Nonetheless, enhancing a system's cybersecurity posture early in the development and production stages lowers the attack surface and, as a result, the cyber threats. The need to develop and implement cybersecurity by design principles for the entire space infrastructure stems from the growing use of COTS components, the commercialization of these sectors, and the growing reliance on software applications. More investigation is required to offer practical guidance on this.

Given the cutting-edge cybersecurity techniques examined in the sections above, a significant unresolved issue with satellite communication systems is striking a delicate balance between security and efficiency ~\cite{mao2023security}.  Although the efficiency of Satcoms protocols is prioritized by their intrinsic design, which reduces power consumption, memory usage, and transmission latency, the implementation of strong security measures may result in significant overhead that isn't always in compliance with mission requirements ~\cite{su2019broadband}. Future research and development ought to develop lightweight security solutions that seamlessly interact with mission requirements in order to meet this challenge.  Among the creative methods are the direct integration of hardware security mechanisms into Satcoms hardware ~\cite{hu2020overview}, the investigation of sophisticated encryption algorithms that provide increased security with negligible overhead, and the creation of adaptive security protocols that can dynamically modify security levels in response to mission requirements. 

The need for autonomous technologies that can function independently of ground control and crew interactions is growing as we get ready to travel farther into space. An increasingly important instrument for achieving this objective is artificial intelligence (AI). A group of Airbus researchers investigated how AI can gather and analyze data aboard the ISS's Columbus module to enhance its prognosis and defect detection skills with assistance from ESA's Discovery program ~\cite{ESA_AI_Space_2025}. The effectiveness of IDP systems using AI is limited by a lack of historical data, restricted collection of data, unknown attack patterns that result in high false positive rates, and limited computing power and memory on spacecraft.  Here, the AI-enabled technique is advantageous, but preserving space AI itself is even more important—keeping an eye out for AI-powered attacks like Deep Locker and Malware-GAN ~\cite{kirat2018deeplocker,rigaki2018bringing} while safeguarding models and data.

Despite the fact that space is a highly regulated field, cybersecurity-specific rules and regulations are inadequate. To strengthen the cybersecurity posture in space, industry standards and recommendations must be adopted; research can significantly aid and inform this process.
\vspace{-0.1in}
\section{Related Work}
The expanding importance of the topic is evidenced by the recent sharp rise in interest in cybersecurity facets of space exploration in both the academic and industrial sectors.The authors in~\cite{pavur2020sok} offer a cross‑disciplinary “threat matrix toolbox” and an original 60‑year chronology of over 100 satellite hacking incidents, then assesses the state‑of‑the‑art across four sub‑domains (radio‑link, hardware, ground station, mission) to chart future research directions. The work in~\cite{peled2023evaluatingsecuritysatellitesystems} presents a comprehensive taxonomy of adversarial tactics, techniques, and procedures against LEO satellites—extending MITRE ATT\&CK to the space domain and illustrates it with case studies including the Viasat outage in Ukraine and the ICARUS DDoS attack. On the other hand,~\cite{299886} systematically investigates the integrity and revocation pitfalls of satellite PKI under orbital delays, and user‑to‑satellite signal‑based location‑privacy risks, identifying research gaps to guide future secure space‑network designs. The work in ~\cite{willbold2023space} emphasizes security of satellite firmware by presenting a taxonomy of risks to satellite firmware and analyzing three real-world satellite firmware pictures experimentally. The experimental vulnerability assessment's findings demonstrate that contemporary in-orbit satellites frequently lack reliable access safeguards and have various software security flaws. Compared to prior works, this paper explicitly organizes and compares threats, impacts, and recovery characteristics across all four space‑infrastructure segments—Ground Station, Space Station, Satellite, and Constellations that contrast traditional vs. space‑specific cybersecurity challenges, attack vectors, and severity metrics using structured tables. By coupling a quantitative risk‑scoring model with a deep dive into mitigation gaps, this SoK bridges descriptive threat cataloging and actionable, prioritized risk management in ways the earlier SoK papers did not.
\section{Conclusion}
The importance of space assets is growing in the interconnected world of today. Since cyberattacks on space systems can have serious repercussions, ranging from communication loss to revealing sensitive information, this domain's cybersecurity has become a major worry as our reliance on satellite technology grows. For this survey, we have thoroughly examined the corresponding cyberattacks and cybersecurity strategies for the four main space system segments—the ground segment, the space segment, the satellite segment and the satellite constellations segment. We have created taxonomy schemes and a Risk Scoring mechanism for the cyberattacks unique to each segment. Given that cyberattacks have the potential to disrupt communication services, compromise private information, physically harm satellites, jam, and spoof GPS signals, and even launch cyber warfare, the consequences underscore how crucial cybersecurity is for this sector.
Additionally, we have included the primary unresolved issues that still exist in this field, along with the relevant directions for further study. As a result, this paper offers a thorough understanding of how cybersecurity in space infrastructure is evolving.

\bibliographystyle{abbrv}
\bibliography{sample}

\end{document}